\documentclass[superscriptaddress, preprint, amsmath,amssymb, aps, pra]{revtex4-1}

\usepackage{graphicx}
\usepackage{dcolumn}
\usepackage{bm}

\begin{document}

\title{Suspense and Surprise in the Book of Technology: Understanding Innovation Dynamics}

\author{Oh-Hyun Kwon}
\affiliation{Pohang University of Science and Technology, Pohang, Republic of Korea}

\author{Jisung Yoon}
\affiliation{KDI School of Public Policy and Management, Sejong, Republic of Korea}

\author{Lav R. Varshney}
\affiliation{University of Illinois Urbana-Champaign, Champaign, IL, USA}

\author{Woo-Sung Jung}
\affiliation{Pohang University of Science and Technology, Pohang, Republic of Korea}

\author{Hyejin Youn}
\email{h.youn@snu.ac.kr}
\affiliation{Graduate School of Business, Seoul National University, Seoul, Republic of Korea}
\affiliation{Santa Fe Institute, Santa Fe, NM, USA}
\affiliation{Kellogg School of Management, Northwestern University, Evanston, IL, USA}
\affiliation{Northwestern Institute of Complex Systems, Evanston, IL, USA}


\begin{abstract}
We envision future technologies through science fiction, strategic planning, or academic research. Yet, our expectations do not always match with what actually unfolds,  much like navigating a story where some events align with expectations while others surprise us. This gap indicates the inherent uncertainty of innovation—how technologies emerge and evolve in unpredictable ways. Here, we elaborate on this inherent uncertainty of innovation in the way technologies emerge and evolve. We define suspense captures accumulated uncertainty and describing events anticipated before their realization, while surprise represents a dramatic shift in understanding when an event occurs unexpectedly. We identify those connections in U.S. patents and show that suspenseful innovations tend to integrate more smoothly into society, achieving higher citations and market value. In contrast, surprising innovations, though often disruptive and groundbreaking, face challenges in adoption due to their extreme novelty. We further show that these categories allow us to identify distinct stages of technology life cycles, suggesting a way to identify the systematic trajectory of technologies and anticipate their future paths.


\end{abstract}


\maketitle

\section{Introduction}
Technological innovation profoundly impacts our daily lives, introducing new products and technologies that shape society. It drives positive economic outcomes, enhancing firm performance \cite{Singhal2020Technological} and influencing product markets \cite{Kock2011The, kogan2017technological}. However, innovation also brings side effects, such as increasing inequality and enabling surveillance of the public \cite{Mao2020Technology}. Therefore, predicting future innovation and preparing for the changes become crucial for the market, policy, and society. Then, how can we predict future innovations?

New knowledge and technologies often emerge from the combination of existing ideas. As Newton famously remarked, ``we stand on the shoulders of giants," technological progress arises through the novel integration of established technologies. For example, Thomas Edison combined carbon filaments, inert gas, and electricity for the first time to invent the incandescent light bulb. Predicting such innovations can be framed as a link prediction task on a network of technological components. Then, the question becomes more specific: can we predict the links that signify future innovations?

Many models are designed to predict innovations for purposes such as strategic planning, policy development, societal preparedness, and academic understanding of the innovation process. Some studies forecast future technologies by identifying universal patterns, like Moore’s Law \cite{farmer_2016_how}, while others investigate technological pathways using life cycle models focused on products and markets \cite{Huang2020Exploring, Markard2020The}. More recently, advancements in machine learning have enabled researchers to extract latent knowledge from empirical data, offering new insights into the dynamics of technological innovation \cite{blount_2012_genomic, tshitoyab2019unsupervised, weis2021learning, lee_2022_technology, gottschalk2023predicting}. 

Yet, a significant challenge in predicting innovation lies in the inherent uncertainty that accompanies technological progress. Many breakthroughs arise from unconventional connections between technologies that are unpredictable by nature. Although such combinations often lead to failures, they also expand the technological space and create opportunities for transformative advancements \cite{rosenburg1996uncertainty, fleming2001recombinant, feng2023surprising}. Existing studies have attempted to measure uncertainty, but typically focus on its decline before innovation occurs \cite{Wang2008Evaluating}. This approach overlooks the dynamic role of uncertainty in generating unexpected breakthroughs. Therefore, a more integrated framework is needed—one that combines predictive models with a systematic understanding of uncertainty—to capture the complex interplay between innovation and the broader societal and technological context.

We propose broadening the perspective on innovation prediction by framing it as an information-revealing process where society interacts with the technological space to forecast the future. In this context, society, much like an audience, formulates predictions based on the information revealed by past technologies. The misalignment between these predictions and realizations gives rise to the suspense and surprise in society \cite{ely2015suspense}.

Suspense and surprise describe how uncertainty accumulates and how expectations deviate from reality. Suspense reflects the cumulation of uncertainty over time, like in a detective novel where the mystery remains unresolved until the climax. Conversely, surprise captures a radical shift in beliefs, much like an unexpected plot twist in a story. This analogy applies to technology prediction as well. For example, in the 1989 movie Back to the Future Part II, filmmakers accurately predicted technologies like virtual meetings in 2015. However, unrealized technology, such as a flying car, creates suspense, which is prolonged anticipation awaiting realization. If flying cars eventually become a reality, they would classified as a suspense innovation. On the other hand, the unexpected rise of artificial intelligence, which the filmmakers did not foresee, would be classified as a surprise innovation.

Using patent data labeled with classification codes representing technological domains, we can systematically categorize suspense and surprise in innovation. Patent data provides an ideal resource for studying technology combinations due to its extensive temporal coverage \cite{Youn2015, Kim2016, lobo2019}. Revisiting Edison’s incandescent light bulb, the patent can be categorized into four distinct technological classes: {\it Composition} (carbon material), {\it Plastic and Nonmetallic Article Shaping} (filament), {\it Distillation} (inert gas), and {\it Electric Lamp} (electricity). Then, we can represent the innovation of the incandescent light bulb with combinations of those technology codes.

In this paper, we classify technological innovations into suspense and surprise using U.S. patent data from 1840 to 2015. We define suspense innovations as connections that were predicted before their realization, whereas surprise innovations are those that were entirely unexpected. We analyze the historical context of these categories through case studies of Edison and Tesla and examine the impacts of suspense and surprise on the technology system. Secondly, we explore the life cycles of various technologies, revealing that cutting-edge technologies often follow technology life cycle patterns, while conventional technologies tend to have off-cycle patterns with more reflection of their foundational role as tools.

\section{Data and Method}
\begin{figure*}[ht]
    \centering
    \includegraphics[width=\textwidth] {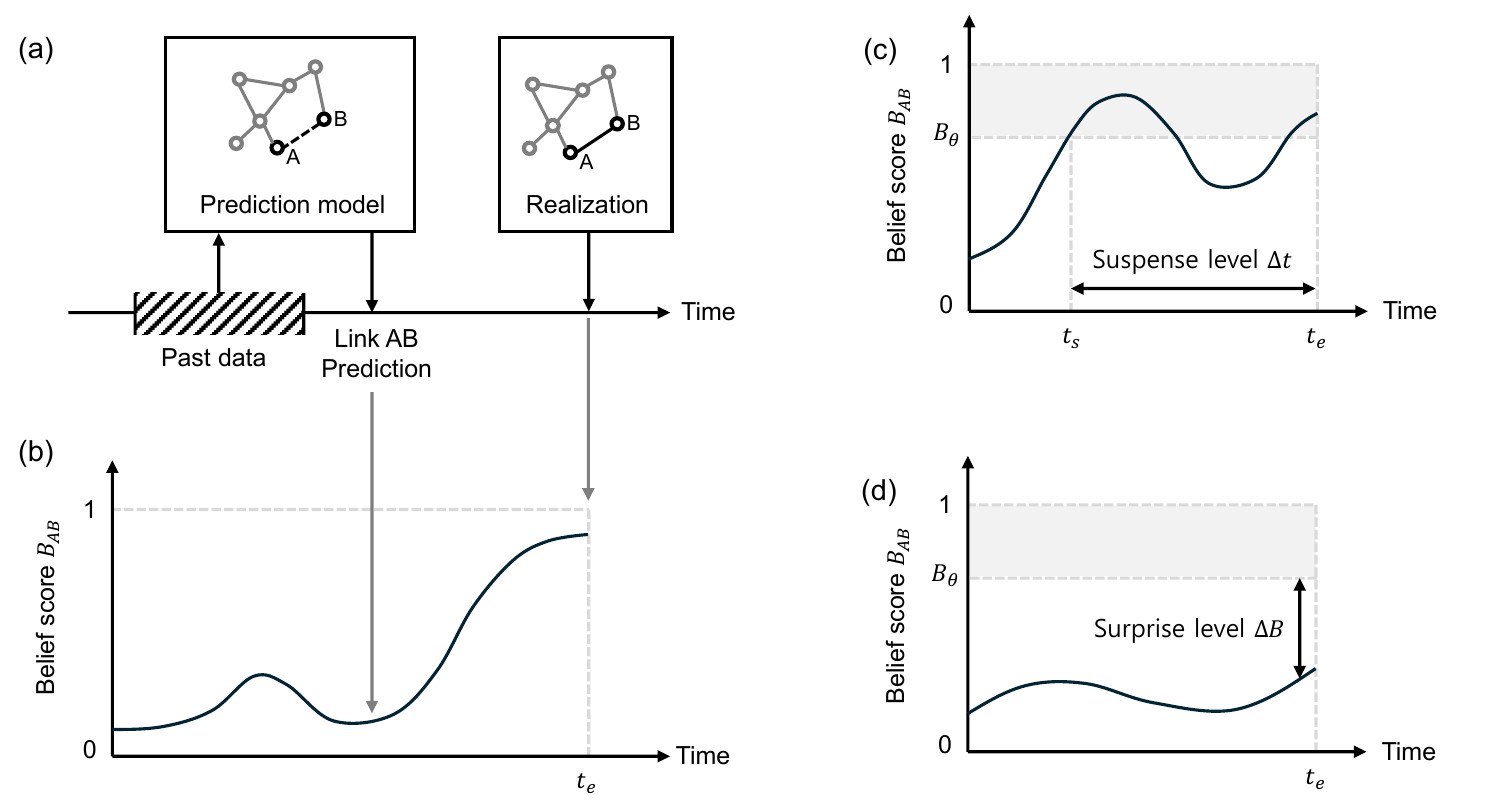}
    \caption{{\bf Suspense and Surprise in Link Prediction.} (a) Using historical data, the prediction model estimates the likelihood of a new connection between technologies A and B, forecasting future events of realization. (b) The model generates a time series of belief scores, indicating how probable it is for the link between A and B to be established. The definition of (c) suspense and (d) surprise. $B_{\theta}$ represents the threshold of prediction, which represents the level of confidence. Note that suspense forms at $t_s$ and the event is realized at $t_e$.}
    \label{fig:method}
\end{figure*}

Figure \ref{fig:method} illustrates the framework of suspense and surprise in technological innovation. We first define belief score $B_{AB}(t)$, which describes the likelihood of a connection between technology A and B at time $t$. This belief score relies on a prediction model that generates possible events given the network configuration, such as link AB in Fig. \ref{fig:method}a. The belief score changes over time since the past events are updated over time. These scores are then compared with observed realization events to categorize them as either {\it suspense} or {\it surprise} events.

\subsection{Data}
We apply the empirical framework to the United States Patent and Trademark Office (USPTO) patent dataset, which provides a comprehensive record of patents issued in the United States. Each patent is assigned technology codes that classify its technological domain, specifically the United States Patent Classification (USPC) codes at the 2-digit main class level. For example, code 977 represents {\it Nanotechnology}. The dataset includes 8,884,966 utility patents spanning from 1940 to 2015, with 438 technology codes. We limit our analysis to patents issued by 2015 because the USPC system was replaced by the Cooperative Patent Classification (CPC) system thereafter. In our study, we define the appearance of a technology linkage based on the patent's issue year, which serves as the temporal reference point for analyzing the emergence of new connections. This approach enables us to investigate how novel combinations of existing technologies drive innovation. With the patent data, we trace the technological innovation over time by investigating the new connections between classification codes.



\subsection{Prediction model}
Along with various options for prediction models, we introduce a simple and explainable approach that captures the general logic of human expectation (the prediction model in Fig \ref{fig:method}a). Our prediction model accounts for two important factors in forecasting future innovation: popularity and similarity. The predicted number of connections between between codes $A$ and $B$, $\hat{N}_{AB}$, is expressed as 
\begin{equation}
    \hat{N}_{AB} = N_AN_B \exp{(c_0+ c_1S_{AB})}.
\end{equation}
Here, $N_A$ and $N_B$ represent the number of patents containing code $A$ and $B$, respectively, and thus $N_AN_B$ indicate their popularity. The idea is that codes frequently used in patents are more likely to make connections. The second term $S_{AB}$ captures the context similarity between codes $A$ and $B$ \cite{tacchella2020innovation}, indicating how often the two codes are used in similar contexts. For example, codes such as {\it Telegraphy} and {\it Television} exhibit high similarity because they are frequently used in a similar technological context. We calculate $S_{AB}$ using a word2vec model \cite{mikolov2013distributed}, treating patents as sentences and technology codes as words. To account for temporal trends, we apply a 5-year moving window to generate context similarity scores. To minimize the influence of word order, we randomize the sequence of technology codes in each patent and average the cosine similarity across 5 iterations. Note that we can calculate the similarity between codes even if they have never appeared together in a patent.

To calibrate the model, we use realized connections between codes, $N_{AB}$, and fit a regression model $\log(N_{AB}/N_AN_B) = c_1 S_{AB} + c_0$. Using the estimated coefficients, we compute the expected number of connections $\hat{N}_{AB}$ and normalize it annually with the cumulative distribution function (CDF) to obtain a belief score $B_{AB}(t)$. The belief score ranges from 0 to 1, where the code pair with the highest $\hat{N}_{AB}$ receives a value of 1. This normalization provides a time series of belief scores for every possible code pair, as illustrated in Fig. \ref{fig:method}b.

\subsection{Suspense and surprise}
We define the concepts of suspense and surprise \cite{ely2015suspense} within the context of technological innovation, adapting their original definitions from entertainment narratives. Unlike entertainment scenarios, which explore multiple possibilities shaped by sports rules or an author’s imagination, our data provide only a single realization for each event, without ensembles or alternative outcomes. The belief scores $B_{AB}(t)$ reflects the likelihood that an unrealized pair of technology codes $A$ and $B$ will eventually connect. 

We establish a threshold $B_{\theta}$ to represent the confidence level of the event in the prediction model. If the belief score exceeds this threshold, the model predicts that the link will form in the following year. Conversely, if the belief score remains below the threshold, the model does not expect a connection. We set $B_\theta = 0.68$, which is the optimal threshold determined through Receiver Operating Characteristic (ROC) analysis. Since this optimal value remains stable over time, we use a singular threshold value for the entire time span of our study.

Suspense is defined as the accumulation of uncertainty over time. A new link is classified as a suspense event if the belief score $B_{AB}(t)$ surpasses the threshold $B_{\theta}$ at least once before the connection occurs (Fig. \ref{fig:method}c). When this threshold is crossed, the model indicates that the event could happen at any moment, with uncertainty building until the connection is realized. The level of suspense is quantified by the duration between the first time the belief score exceeds the threshold and the actual connection. This duration is expressed as $\Delta t = t_e - t_s$, where $t_s = \min\{t|B_{AB}(t) > B_{\theta}\}$ is the very first year when the belief score exceeds the threshold, and $t_e$ is the year the connection is realized. With the suspense level, the suspense event is characterized by $\Delta t > 0$, indicating that the model recognized the potential event before it occurred.

Surprise captures a sudden change in beliefs. A new link is classified as a surprise event if the belief score $B_{AB}(t)$ remains below the threshold $B_{\theta}$ in the year before the connection (Fig. \ref{fig:method}d). This indicates that the model did not expect the event prior to its realization. The level of surprise is quantified by the gap between the belief score in the year preceding the connection and the threshold, defined as $\Delta B_{AB} = B_\theta - B_{AB}(t_e-1)$. A surprising event occurs when $\Delta B > 0$, signaling that the belief score was insufficient to predict the connection.

\section{Result}
\subsection{Suspense and Surprise Innovation}
\begin{figure*}[ht]
    \centering
    \includegraphics[width=0.8\textwidth]{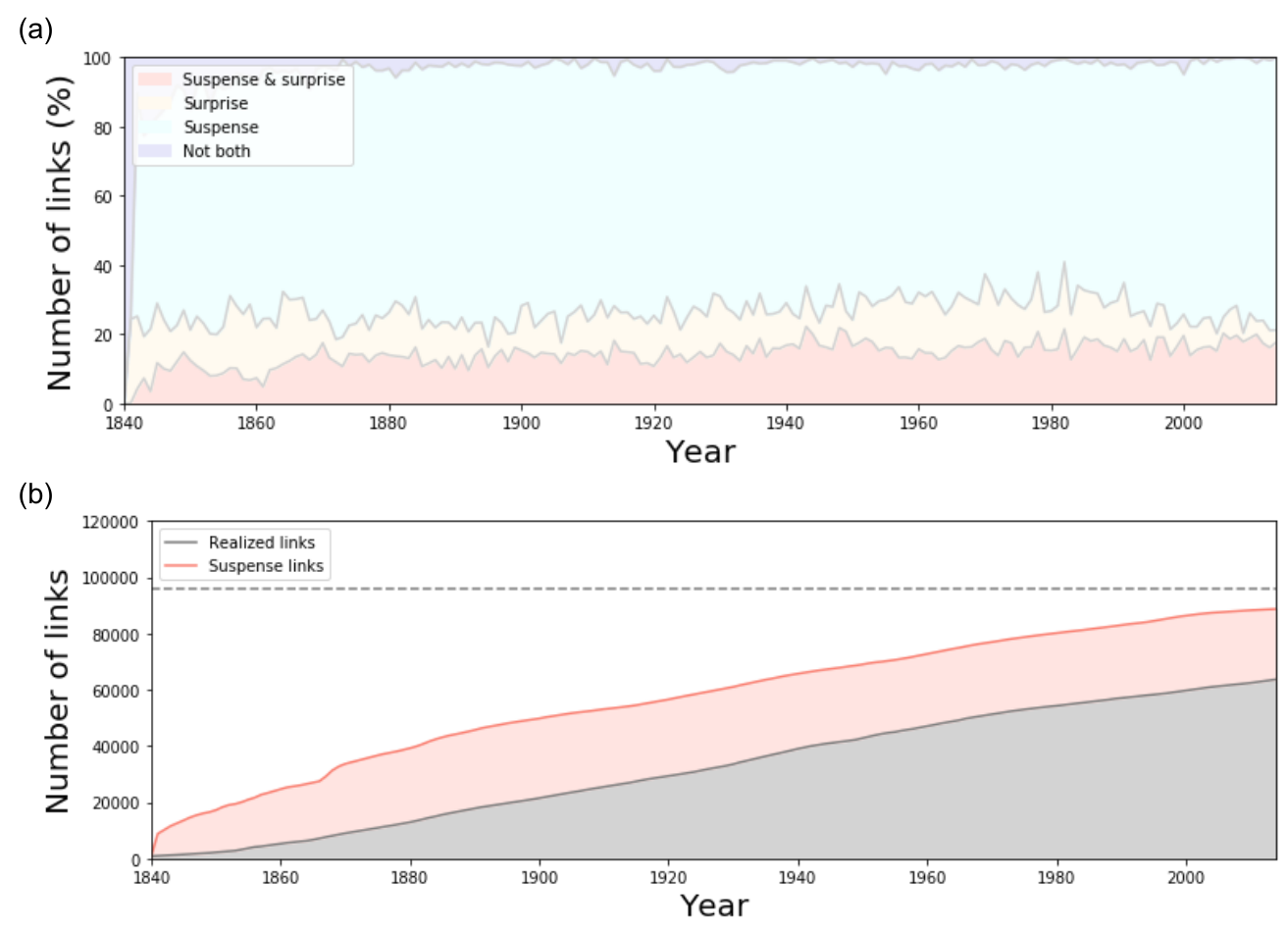}
    \caption{{\bf Suspense and Surprise Over Time.} (a) The fraction of events categorized into four groups: timely prediction, suspense, surprise, and both suspense and surprise. (b) The number of realized connections and unrealized suspense (pairs predicted to connect but not yet connected) over time. The dotted line indicates the possible maximum number of connections. The number of unrealized suspense has remained relatively stable, hovering around 20,000.}
    \label{fig:result1}
\end{figure*}

We analyze US patents to classify connection events at time $t$  into four categories: timely prediction, suspense, surprise, and both suspense and surprise, as shown in Fig. \ref{fig:result1}a. Our model identifies 95,703 potential links, of which 63,696 were connected by 2015. These 63,696 comprise 53,533 suspenseful connections (connections that were anticipated before they were made) and 16,723 as surprising (model failed to predict). Notably, 9,131 connections met the criteria for both suspense and surprise, which makes them particularly interesting. 

These dual-classified events are possible because our categories have two different dimensions for prediction: timing and actual connections. When connections are predicted, they are considered suspenseful because we anticipate ($B$) their likelihood of occurring soon, making them unlikely to surprise us. However, when these anticipated events fail to materialize over an extended period, our belief factor ($B$) diminishes. If they eventually occur after being largely forgotten, they simultaneously fulfill the criteria for suspense and surprise. In other words, these connections represent combinations that were initially promising but lost prominence over time, only to emerge unexpectedly later. Figure \ref{fig:result1}a illustrates the fraction of suspenseful and surprising events over time, showing that most new connections fall into one or both categories, with the ratio between suspense and surprise remaining relatively stable throughout history.

We find that suspense innovations are often associated with fundamental technologies, and surprise innovations connect cutting-edge technologies. Suspense events connect technologies in mechanical systems, where connections are easier to predict due to their broad applicability and established domains. In contrast, surprise innovations are more likely to involve cutting-edge technologies, such as nanotechnology or artificial intelligence. For instance, connections around fundamental mechanics are more predictable, while predicting connections for advanced technologies like the intersection of artificial intelligence and neuroscience (e.g., neural networks) is considerably more challenging. Interestingly, some technologies exhibit both suspense and surprise characteristics. For example, tools such as brushes, whips, knots, and ammunition demonstrate connections that are both anticipated and unexpected, reflecting their adaptability and evolving nature.

Inventors continually draw from this pile of ideas to bring connections to realization. Figure \ref{fig:result1}b highlights the number of realized links and unrealized suspense (pairs that exceeded the threshold $B_{\theta}$ but have not yet connected) over time. The number of unrealized suspense pairs has remained relatively consistent, at around 20,000, throughout the observed period. Those unrealized suspense pairs emerge and are eventually realized, suggesting a relatively stable reservoir of ``ideas in the air."

\subsection{What Do Suspense and Surprise in Innovation Tell Us?}
Suspenseful connections generate more attention and profits when they are realized, while the impact of surprising connections is harder to predict. Suspenseful connections are generally well-accepted because they are more conventional and anticipated, making it easier for society to integrate the resulting technology. In contrast, surprising connections often have less impact on average and deliver more unpredictable outcomes, but they occasionally lead to groundbreaking technological breakthroughs.

\begin{figure*}[ht]
    \centering
    \includegraphics[width=0.9\textwidth]{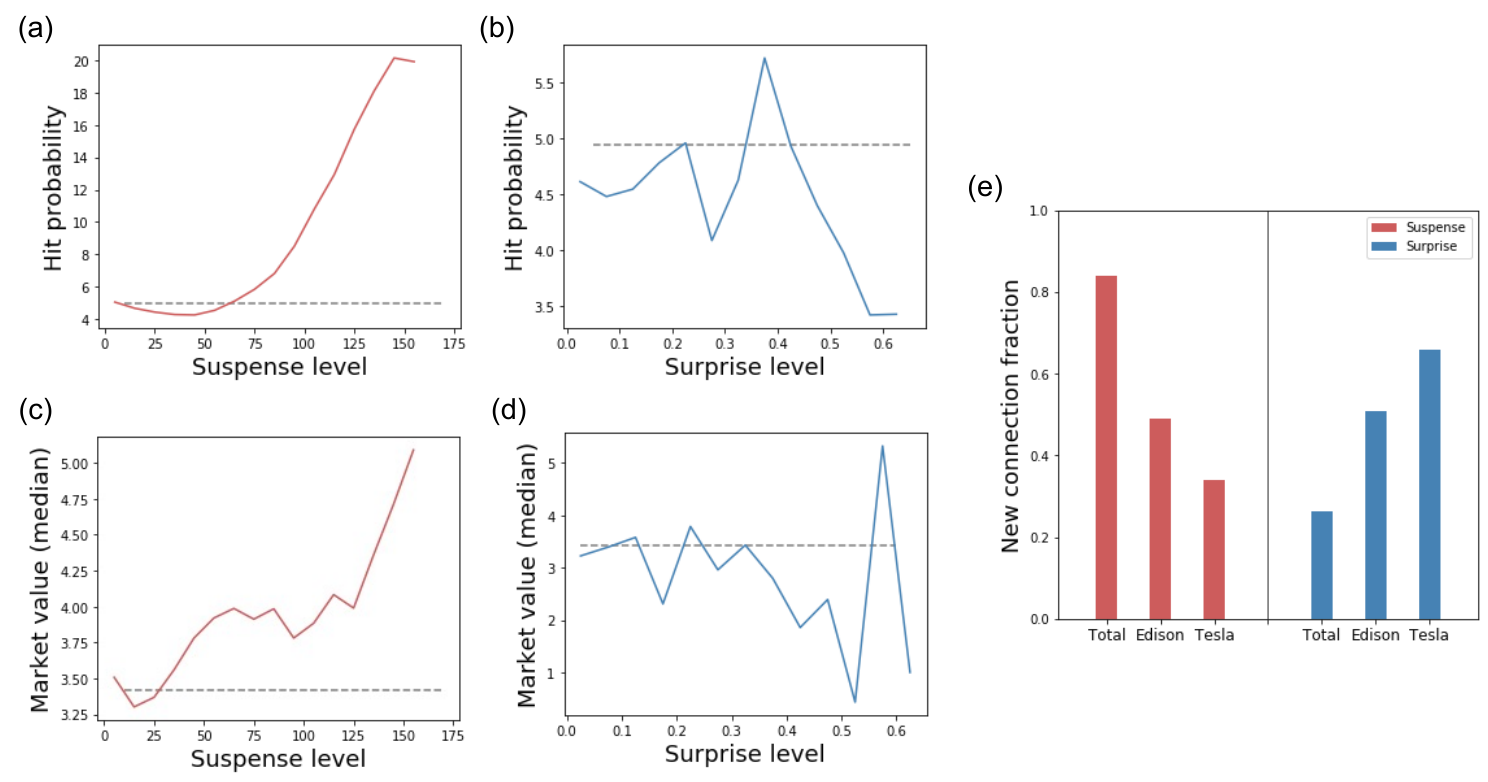}
    \caption{{\bf Impact by Suspense and Surprise.} (a-b) Top 5\% 10-year citation hit probability for different levels of (a) suspense and (b) surprise. The dotted line describes the hit probability of the entire dataset. (c-d) Median of market value for different levels of (c) suspense and (d) surprise. The dotted line represents the median market value of the entire dataset. (e) The fraction of suspense and surprise events in connections by Edison and Tesla compared to the total dataset. Edison and Tesla both exhibit less suspense and more surprise connections than total dataset. However, Edison shows a balanced fraction, while Tesla has an extreme fraction.}
    \label{fig:result2}
\end{figure*}

We evaluate the impact of suspense and surprise connections using two metrics that represent the attention and profitability of the technology. The first metric is hit probability, which measures the likelihood that a connection ranks in the top 5\% of citations within 10 years. The second metric is market value, reflecting the economic impact indicated by market changes following the release of the patent. Since citation and market value data only cover a subset of the dataset, the analysis is limited to the available data.

Technologies with suspenseful connections tend to gain more impact, particularly when the connection has been anticipated for a long period. The hit probability for suspenseful connections is higher than the average for the entire dataset and increases with the level of suspense (Fig. \ref{fig:result2}a). This shows that connections with prolonged anticipation are more likely to achieve high citation counts, highlighting their importance. Similarly, the market value of suspenseful connections is higher and increases with the level of suspense (Fig. \ref{fig:result2}c).

In contrast, overly disruptive technologies often lose impact. The hit probability for surprise connections fluctuates below the average for the dataset (Fig. \ref{fig:result2}b). As the level of surprise increases, the hit probability decreases, suggesting that extremely surprising connections are less likely to be recognized or adopted. Similarly, the market value of surprise connections is also lower and fluctuates below the average for the dataset (Fig. \ref{fig:result2}d).

However, these results do not imply that suspenseful connections are better than surprising ones, or vice versa. In fact, balancing convention and novelty is crucial when combining ideas \cite{uzzi2013atypical}. To explore this balance, we examine the inventions of Edison and Tesla during the era of electricity. Both inventors contributed significantly to society, but with different approaches. Edison focused on business-oriented inventions, prioritizing practical applications that could be integrated into society. Tesla, on the other hand, concentrated on experimental ideas that laid the groundwork for future technological breakthroughs.

By comparing their patents, we find that Edison created 153 new links across 1,084 patents, while Tesla generated 50 new links from 112 patents. Edison’s innovations included 75 suspense pairs and 78 surprise pairs, whereas Tesla produced 17 suspense pairs and 33 surprise pairs. Proportionally, Tesla generated a higher share of surprising pairs compared to Edison (Fig. \ref{fig:result2}e).

Interestingly, both inventors created fewer suspenseful and more surprising connections than observed in the overall dataset. Edison’s innovations struck a balance between average and extreme connections, reflecting his focus on institutionalized inventions with broad societal impact \cite{yang2024geometrics}. In contrast, Tesla’s work leaned toward pure engineering breakthroughs, with a strong emphasis on surprising innovations that pushed the boundaries of existing technologies.

These findings highlight the roles of suspense and surprise in innovation. Suspenseful connections, with their higher anticipation and integration potential, are more likely to gain attention and economic success, making them a valuable part of the technological ecosystem. Surprising connections, while groundbreaking, often face challenges in achieving widespread acceptance. Then, how do suspense and surprise find balance? From a global perspective, suspense and surprise tend to alternate throughout the life cycle of technology.

\subsection{Suspense and Surprise in Technology Life Cycle}
\begin{figure*}[ht]
    \centering
    \includegraphics[width=0.9\textwidth]{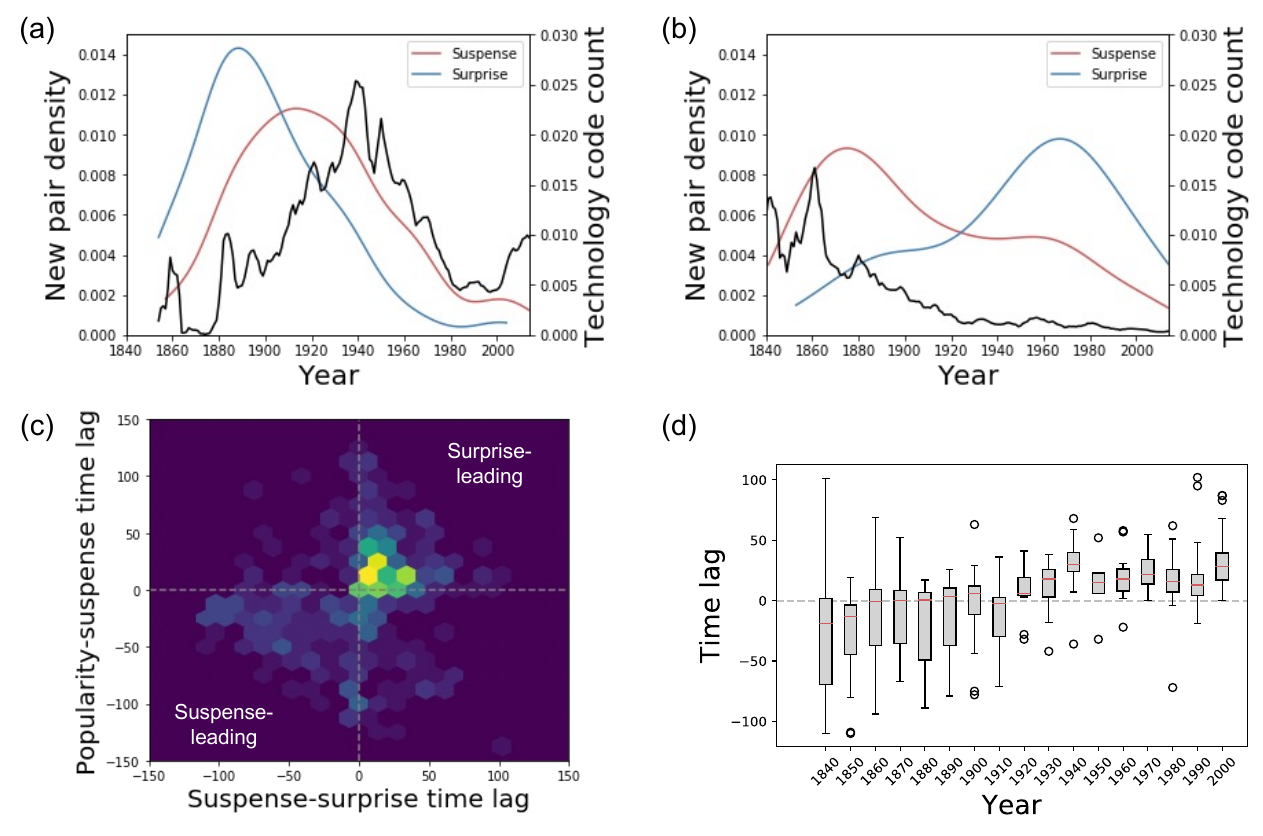}
    \caption{{\bf Life Cycle of Technology Through Surprise, Suspense, and Popularity.} (a-b) The technology lifecycle is shown for (a) {\it Electric Lamp and Discharge Devices} and (b) {\it Planting}. The red bars and blue curves represent new connections classified as suspense and surprise, respectively, with curves estimated using kernel density estimation (KDE) to track the number of suspense and surprise events over time. The black line illustrates the usage of the technology code (popularity) throughout the observed period. (c) The distribution of time lags shows the relationship between suspense and surprise. The upper-right quadrant corresponds to on-cycle innovations (surprise - suspense - popularity), while the lower-left quadrant represents off-cycle innovations (popularity - suspense - surprise). (d) Temporal changes in time lag are displayed. Positive time lags indicate on-cycle innovations, and the majority of innovations since 1920 have followed on-cycle patterns.}
    \label{fig:result3}
\end{figure*}

Suspense and surprise highlight the stages of the technology life cycle. The technology life cycle captures the market response to technological development, progressing through phases of innovation, syndication, diffusion, and substitution. To illustrate these phases, we aggregate connections at the technology level, encompassing all connections associated with a given technology. Figures \ref{fig:result3}a and \ref{fig:result3}b depict these trends for the technical codes {\it Electric Lamp and Discharge Devices} and {\it Planting}. In these figures, the blue and red lines represent the trends of suspense and surprise, respectively, estimated using kernel density estimation (KDE).

The majority of technologies follow the well-known technology life cycle. In this typical pattern, technologies progress through a sequence of surprise, suspense, and popularity peaks (Fig. \ref{fig:result3}a). This lifecycle reflects a widely recognized model of the technology life cycle: innovation (surprise peak), syndication (suspense peak), diffusion (popularity peak), and substitution (decline in popularity). Approximately 44.74\% of new connections follow this on-cycle pattern (Fig. \ref{fig:result3}c, upper-right quadrant), with most examples emerging after 1920 (Fig. \ref{fig:result3}d). Cutting-edge technologies frequently exhibit this life cycle.

However, some technologies deviate from this cycle, exhibiting off-cycle patterns. These often include older or foundational technologies that are less focused on and do not follow the traditional life cycle. A common off-cycle pattern begins with an initial popularity peak, followed by suspense and then surprise peaks (Fig. \ref{fig:result3}b). This pattern is typical of technologies where usefulness evolves over time. Inventors initially concentrate on practical and incremental improvements (suspense pairs) before exploring more unconventional or groundbreaking ideas (surprise pairs). This pattern accounts for 22.62\% of new connections (Fig. \ref{fig:result3}c, lower-left quadrant) and was more common before 1920 (Fig. \ref{fig:result3}d). Foundational technologies often display this off-cycle behavior. 

Notably, there are still remaining 23.97\% off-cycle patterns that are hard to categorize. These cases may indicate early signs of upcoming life cycles, reflecting the diverse and complex nature of technological development. The transition in dominant lifecycle patterns around 1920 may be attributed to historical events such as the World Wars or changes in patent policies.



\section{Discussion}

In this paper, we categorize innovation into suspense and surprise events from the perspective of the information-revealing process. We use U.S. patent data spanning 1840 to 2015 to categorize suspense and surprise in technological innovation. We examine the historical context of these categories through case studies of Edison and Tesla and evaluate their impacts on the technological ecosystem. Additionally, we investigate the life cycles of various technologies, showing that cutting-edge innovations typically align with traditional life cycle patterns, whereas conventional technologies often exhibit off-cycle patterns that reflect their foundational role as tools.

One application of our framework is the prediction of upcoming trends in technologies. For instance, we track the ups and downs of technology such as artificial intelligence, or we can monitor which technology is at which phase of the life cycle.

Our framework can be extended to any system where the misalignment between prediction and reality plays a critical role. First, it can be applied to other domains with different datasets, such as predicting the emergence of new knowledge in academia or identifying dramatic plot twists in movies. Second, the framework allows for integration with alternative prediction models. For instance, machine learning algorithms could be used to study how AI-driven predictions reshape the information-revealing process. Alternatively, models with specific mechanisms, such as focusing solely on context similarity, could be developed. Finally, the link prediction task itself can be expanded to include hyperlinks, representing higher-order combinations of technologies to capture more complex innovation patterns.

Our approach also has some limitations. First, our model is restricted to detecting links between existing technologies within the classification system. It cannot identify combinations involving technologies that fall outside the defined classification codes. Additionally, the predictions rely on the structure of knowledge embedded in the classification system, which is partially influenced by the subjective judgment of individuals assigning the codes. Another limitation is that the dataset only includes patents from the United States, potentially limiting the global applicability of the findings. 


\bibliographystyle{ieeetr}
\bibliography{reference.bib}

\begin{thebibliography}{10}

\bibitem{Singhal2020Technological}
C.~Singhal, R.~V. Mahto, and S.~Kraus, ``Technological innovation, firm performance, and institutional context: A meta-analysis,'' {\em IEEE Transactions on Engineering Management}, vol.~PP, pp.~1--11, 2020.

\bibitem{Kock2011The}
A.~Kock, H.~Gemünden, S.~Salomo, and C.~Schultz, ``The mixed blessings of technological innovativeness for the commercial success of new products,'' {\em Journal of Product Innovation Management}, vol.~28, pp.~28--43, 2011.

\bibitem{kogan2017technological}
L.~Kogan, D.~Papanikolaou, A.~Seru, and N.~Stoffman, ``Technological innovation, resource allocation, and growth,'' {\em The Quarterly Journal of Economics}, vol.~132, p.~665–712, 5 2017.

\bibitem{Mao2020Technology}
C.~Mao, R.~Koide, A.~Brem, and L.~Akenji, ``Technology foresight for social good: Social implications of technological innovation by 2050 from a global expert survey,'' {\em Technological Forecasting and Social Change}, vol.~153, p.~119914, 2020.

\bibitem{farmer_2016_how}
J.~D. Farmer and F.~Lafond, ``How predictable is technological progress?,'' {\em Research Policy}, vol.~45, pp.~647--665, 04 2016.

\bibitem{Huang2020Exploring}
Y.~Huang, F.~Zhu, A.~Porter, Y.~Zhang, D.~Zhu, and Y.~Guo, ``Exploring technology evolution pathways to facilitate technology management: From a technology life cycle perspective,'' {\em IEEE Transactions on Engineering Management}, vol.~68, pp.~1347--1359, 2020.

\bibitem{Markard2020The}
J.~Markard, ``The life cycle of technological innovation systems,'' {\em Technological Forecasting and Social Change}, 2020.

\bibitem{blount_2012_genomic}
Z.~D. Blount, J.~E. Barrick, C.~J. Davidson, and R.~E. Lenski, ``Genomic analysis of a key innovation in an experimental escherichia coli population,'' {\em Nature}, vol.~489, pp.~513--518, 09 2012.

\bibitem{tshitoyab2019unsupervised}
V.~Tshitoyan, J.~Dagdelen, L.~Weston, A.~Dunn, Z.~Rong, O.~Kononova, K.~A. Persson, G.~Ceder, and A.~Jain, ``Unsupervised word embeddings capture latent knowledge from materials science literature,'' {\em Nature}, vol.~571, no.~7763, pp.~95--98, 2019.

\bibitem{weis2021learning}
J.~W. Weis and J.~M. Jacobson, ``Learning on knowledge graph dynamics provides an early warning of impactful research,'' {\em Nature Biotechnology}, vol.~39, no.~10, pp.~1300--1307, 2021.

\bibitem{lee_2022_technology}
M.~Lee, S.~Kim, H.~Kim, and J.~Lee, ``Technology opportunity discovery using deep learning-based text mining and a knowledge graph,'' {\em Technological Forecasting and Social Change}, vol.~180, p.~121718, 07 2022.

\bibitem{gottschalk2023predicting}
F.~Gottschalk, B.~Debray, F.~Klaessig, B.~Park, J.-M. Lacome, A.~Vignes, V.~P. Portillo, S.~V{\'a}zquez-Campos, C.~O. Hendren, S.~Lofts, {\em et~al.}, ``Predicting accidental release of engineered nanomaterials to the environment,'' {\em Nature Nanotechnology}, pp.~1--7, 2023.

\bibitem{rosenburg1996uncertainty}
N.~Rosenberg, ``Uncertainty and technological change,'' {\em Conference Series ; [Proceedings]}, pp.~91--125, 02 1996.

\bibitem{fleming2001recombinant}
L.~Fleming, ``Recombinant uncertainty in technological search,'' {\em Management Science}, vol.~47, no.~1, pp.~117--132, 2001.

\bibitem{feng2023surprising}
F.~Shi and J.~Evans, ``{Surprising combinations of research contents and contexts are related to impact and emerge with scientific outsiders from distant disciplines},'' {\em Nature Communications}, vol.~14, p.~1641, 03 2023.

\bibitem{Wang2008Evaluating}
C.-H. Wang, I.~Lu, and C.-B. Chen, ``Evaluating firm technological innovation capability under uncertainty,'' {\em Technovation}, vol.~28, pp.~349--363, 2008.

\bibitem{ely2015suspense}
J.~Ely, A.~Frankel, and E.~Kamenica, ``Suspense and surprise,'' {\em Journal of Political Economy}, vol.~123, no.~1, pp.~215--260, 2015.

\bibitem{Youn2015}
H.~Youn, D.~Strumsky, L.~M. Bettencourt, and J.~Lobo, ``{Invention as a combinatorial process: Evidence from US patents},'' {\em Journal of the Royal Society Interface}, vol.~12, no.~106, p.~20150272, 2015.

\bibitem{Kim2016}
D.~Kim, D.~Cerigo, H.~Jeong, and H.~Youn, ``Technological novelty profile and invention’s future impact,'' {\em EPJ Data Sci.}, vol.~6, no.~8, 2016.

\bibitem{lobo2019}
J.~Lobo and D.~Strumsky, ``Sources of inventive novelty: two patent classification schemas, same story.,'' {\em Scientometrics}, vol.~120, pp.~19--37, 2019.

\bibitem{tacchella2020innovation}
A.~Tacchella, A.~Napoletano, and L.~Pietronero, ``The language of innovation,'' {\em PLoS ONE}, vol.~15, no.~4, p.~e0230107, 2020.

\bibitem{mikolov2013distributed}
T.~Mikolov, I.~Sutskever, K.~Chen, G.~S. Corrado, and J.~Dean, ``Distributed representations of words and phrases and their compositionality,'' {\em Advances in neural information processing systems}, vol.~26, 2013.

\bibitem{uzzi2013atypical}
B.~Uzzi, S.~Mukherjee, M.~Stringer, and B.~Jones, ``Atypical combinations and scientific impact,'' {\em Science}, vol.~342, no.~6157, p.~468–472, 2013.

\bibitem{yang2024geometrics}
S.~Yang and H.~Youn, ``Geometrics of the adjacent possible: Harvesting values at the curvature,'' 2024.

\end{thebibliography}

\newpage

\appendix

\section*{Supplementary}

\renewcommand{\thefigure}{S\arabic{figure}}
\setcounter{figure}{0}


\subsection{Context similarity}
We calculate context similarity using a word2vec model. By treating classification codes as words and patents as sentences, we vectorize the context of classification codes for each year. A 5-year moving window is used to compute the embedding space for the subsequent year. The model is configured with a vector size of 32 and a window size of 10. This embedding space captures how technologies are used in similar contexts.

To quantify similarity, we measure the cosine similarity between classification code vectors. Importantly, this embedding approach enables the calculation of similarity between classification codes that have never been directly connected. To minimize the randomness inherent in the word2vec model, we ensemble the results of five independent runs, averaging the cosine similarity scores across these runs.

\subsection{Prediction model}
\begin{figure*}[ht]
    \centering
    \includegraphics[width=\textwidth]{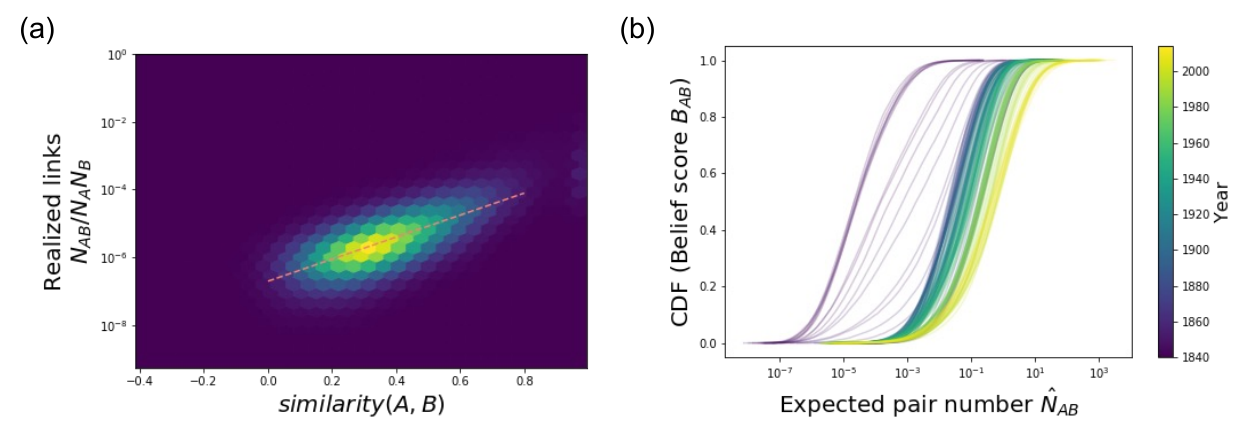}
    \caption{{\bf Prediction model.} (a) Linear relationship between $\log(N_{AB}/N_AN_B)$ and the context similarity. (b) Normalization of the $\hat{N}_{AB}$ using the CDF.}
    \label{fig:predictionmodel}
\end{figure*}

We predict future connections by estimating the expected number of connections between technology codes $A$ and $B$. To calibrate our model, we utilize information from existing links. Specifically, we estimate the coefficients using the regression model $\log(N_{AB}/N_AN_B) = c_1 S_{AB} + c_0$. The regression results are presented in Table \ref{tab:regression} and visualized in Fig. \ref{fig:predictionmodel}a.

\begin{table}[htbp]
    \centering
    \newcommand\sym[1]{\rlap{$^{#1}$}}
    
    \caption{Regression model for $\log(N_{AB}/N_AN_B)$ \label{tab:regression}}

    \begin{tabular}{r|c}
      & Coefficient \\
    \hline
    Similarity & -6.711\sym{***}\\
     & (0.002)\\
    Constant & 3.253\sym{***}\\
     & (0.005)\\
    \hline
    $R^2$ & 0.474\\
    Observations & 2,852,622 \\
    \hline
    \multicolumn{2}{l}{Standard errors in parentheses}\\
    \multicolumn{2}{l}{* p<0.10, ** p<0.05, *** p<0.01}\\
    \end{tabular}
\end{table}

Since the average expected number of connections $\hat{N}_{AB}$ varies over time, normalization is required. To achieve this, we compute the cumulative distribution function (CDF) of $\hat{N}_{AB}$ for each year, mapping the values to a range between 0 and 1 (Fig. \ref{fig:predictionmodel}b). The figure illustrates that the CDF curves shift toward higher values of $\hat{N}_{AB}$ over time, reflecting changes in the data distribution.

\subsection{Belief score threshold $B_{\theta}$ (Level of confidence)}
\begin{figure*}[ht]
    \centering
    \includegraphics[width=0.5\textwidth]{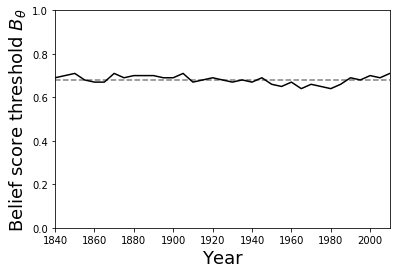}
    \caption{{\bf Belief score threshold by year.} The dotted grey line represent the threshold from total dataset, which is 0.68. The black line represent the belief score threshold $B_{\theta}$ calculated with 5-year moving window data.}
    \label{fig:threshold}
\end{figure*}

In our manuscript, we set the belief score threshold $B_{\theta}=0.68$, identified as the optimal value through Receiver Operating Characteristic (ROC) analysis. Specifically, we determine this threshold by calculating the distance between the point (0, 1) and the ROC curve, selecting the threshold that minimizes this distance. While we use a constant value for $B_{\theta}$ throughout our analysis, we also examine its temporal variation over time (Fig. \ref{fig:threshold}). The results show that the belief score threshold $B_{\theta}$ remains relatively stable over time, consistently approximating the optimal value derived from the complete dataset.

\subsection{Performance of the prediction model}
The prediction model performs well overall, achieving an area under the curve (AUC) of 0.7660 in the ROC analysis. However, the model exhibits lower precision and F1-score due to a high number of false-positive predictions. Many of these false positives correspond to suspense innovations, where connections are anticipated but not immediately realized. While the model effectively suggests multiple possible connections, it struggles to pinpoint the exact timing of innovation, resulting in occasional prediction inaccuracies.

\begin{figure*}[ht]
    \centering
    \includegraphics[width=0.9\textwidth]{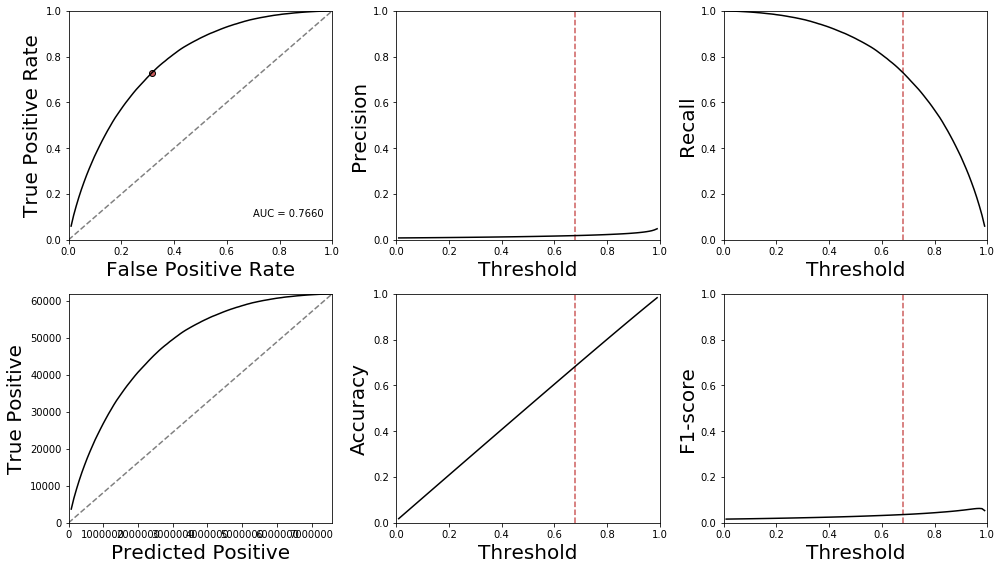}
    \caption{{\bf Performance of the prediction model.} (a) ROC curve, (b) precision, (c) recall, (d) TOC curve, (e) accuracy, and (f) F1 score of the prediction model. The red dotted line represent the optimal threshold on the ROC curve, which is 0.68.}
    \label{fig:prediction model}
\end{figure*}

\subsection{Case study of life cycles}
\begin{figure*}[ht]
    \centering
    \includegraphics[width=0.9\textwidth]{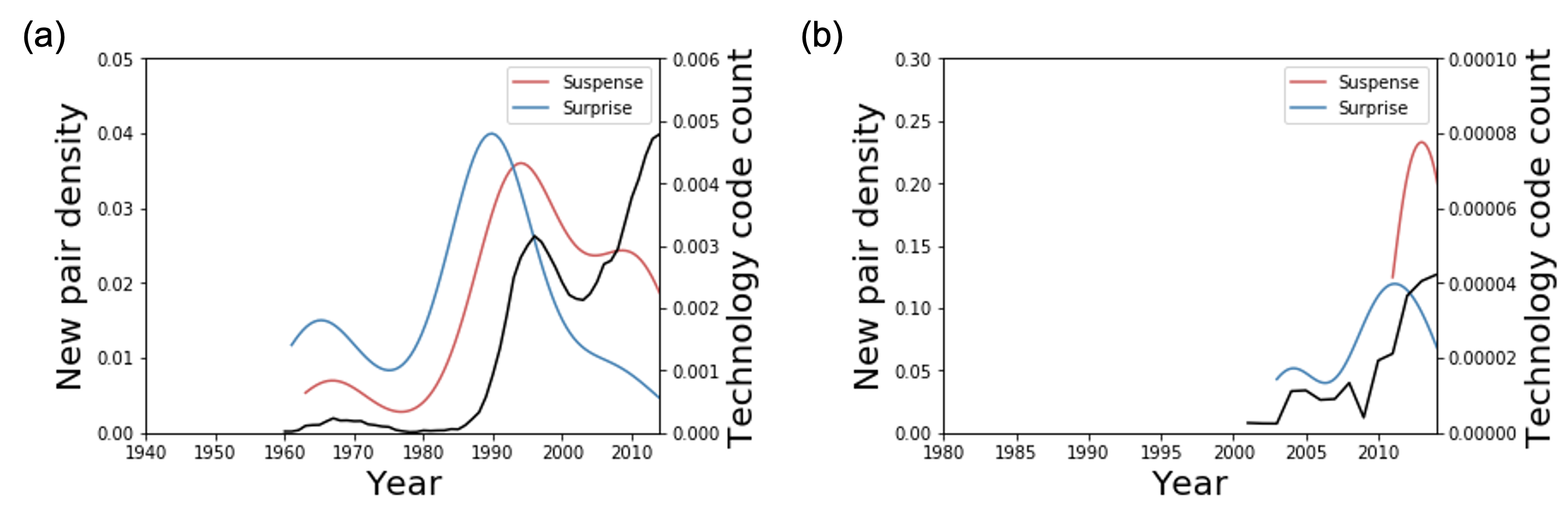}
    \caption{{\bf Case study of technology life cycle.} Life cycle of (a) {\it Data processing: artificial inteligence} and (b) {\it Organic compounds}.}
    \label{fig:example}
\end{figure*}

We can illustrate the insights provided by this analysis of technology life cycles with specific examples. Figure \ref{fig:example}a depicts the life cycle of artificial intelligence (AI), highlighting the alternating booms and winters characteristic of AI development. In another example, Figure \ref{fig:example}b presents the life cycle of organic compounds, which is currently at the suspense peak phase. From this, we can anticipate that organic compound technology will soon transition into the popularity phase as the suspense peak begins to decline.

\end{document}